\documentclass[aps,prl,reprint,superscriptaddress]{revtex4-1}
\usepackage{graphicx}
\usepackage{dcolumn}
\usepackage{bm}
\usepackage{color}
\usepackage{ulem}

\begin{document}


\title{Berry Phase in Cuprate Superconductors}


\author{N. Doiron-Leyraud}
\affiliation{D\'epartement de physique, Universit\'e de Sherbrooke, J1K 2R1 Canada}
\affiliation{Department of Physics, McGill University, Montreal, H3A 2T8 Canada}

\author{T. Szkopek}
\affiliation{Department of Electrical and Computer Engineering, McGill University, Montreal, Quebec, H3A 2A7, Canada}

\author{T. Pereg-Barnea}
\affiliation{Department of Physics, McGill University, Montreal, H3A 2T8 Canada}

\author{C. Proust}
\affiliation{Laboratoire National des Champs Magn\'etiques Intenses,
CNRS, INSA, UJF, UPS, Toulouse, 31400 France}
\affiliation{Canadian Institute for Advanced Research, Toronto, Ontario, Canada M5G 1Z8}

\author{G. Gervais}
\email[]{gervais@physics.mcgill.ca}
\affiliation{Department of Physics, McGill University, Montreal, H3A 2T8 Canada}
\affiliation{Canadian Institute for Advanced Research, Toronto, Ontario, Canada M5G 1Z8}


\date{\today}



\begin{abstract}
The geometrical Berry phase is recognized as having profound implications for the properties of electronic systems. Over the last decade, Berry phase has been essential to our understanding of new materials, including graphene and topological insulators. The Berry phase can be accessed via its contribution to the phase mismatch in quantum oscillation experiments, where electrons accumulate a phase as they traverse closed cyclotron orbits in momentum space. The high-temperature cuprate superconductors are a class of materials where the Berry phase is thus far unknown despite the large body of existing quantum oscillations data. In this report we present a systematic Berry phase analysis of Shubnikov - de Haas measurements on the hole-doped cuprates YBa$_2$Cu$_3$O$_{y}$, YBa$_2$Cu$_4$O$_8$, HgBa$_2$CuO$_{4 + \delta}$, and the electron-doped cuprate  Nd$_{2-x}$Ce$_x$CuO$_4$. For the hole-doped materials, a trivial Berry phase of 0 mod($2\pi$) is systematically observed whereas the electron-doped Nd$_{2-x}$Ce$_x$CuO$_4$ exhibits a significant non-zero Berry phase. These observations set constraints on the nature of the high-field normal state of the cuprates and points towards contrasting behaviour between hole-doped and electron-doped materials. We discuss this difference in light of recent developments related to charge density-wave and broken time-reversal symmetry states.
\end{abstract}
\maketitle

The geometrical Berry phase~\cite{berry_quantal_1984} is widely recognized as having profound implications for the properties of electronic systems~\cite{xiao_berry_2010}. Over the last decade or so, the Berry phase has been essential to our understanding of new materials such as graphene~\cite{zhang_experimental_2005,novoselov_two-dimensional_2005} and topological insulators~\cite{fu_topological_2007,ando_topological_2013}. In general, a non-trivial Berry phase is a result of band crossing as in the case of a massless Dirac point~\cite{mikitik_manifestation_1999}. The Berry phase can be accessed in quantum oscillations (QO) measurements as it contributes to the phase mismatch of electrons in their cyclotron orbits. With their enigmatic pseudogap and superconducting phases, the cuprates are materials where the Berry phase is thus far unknown, despite the availability of QO data. Here, we report the Berry phase in the high-field, normal state of cuprate superconductors. We focus on the enigmatic underdoped regime and, based on QO data, we determine the berry phase contribution to the phase mismatch unambiguously. In the hole-doped materials YBa$_2$Cu$_3$O$_{y}$, YBa$_2$Cu$_4$O$_8$, and HgBa$_2$CuO$_{4 + \delta}$ a trivial Berry phase of 0 mod($2\pi$) is systematically observed, while the electron-doped Nd$_{2-x}$Ce$_x$CuO$_4$ exhibits a significant non-zero Berry phase of 1.4$\pi$. Our results set significant constraints on the microscopic description of the normal state and, in particular, do not support a nodal structure or broken time-reversal symmetry in the hole doped compounds.

As proposed by Berry three decades ago~\cite{berry_quantal_1984}, a Bloch wave function acquires a geometrical phase as its momentum is wound around a closed contour~\cite{fuchs_topological_2010}. In general, a non-zero Berry phase is expected when the contour encloses points of either a band crossing~\cite{mikitik_manifestation_1999}, such as the Dirac points in graphene, or an avoided band crossing, such as the massive Dirac points in boron nitride.  The Berry phase can therefore be associated with coupling between bands and is expected to vanish for isolated bands, such as in the case of degenerate metals.

Semiclassically, electrons in metals undergo cyclotron motion when subjected to a magnetic field.  In momentum space, the magnetic field induces motion in closed contours on the Fermi surface in the plane perpendicular to the applied field.  The requirement that the wave function returns with the same phase after each cycle leads to the Onsager relation, which determines the Landau level (LL) quantization,

\[A_n \frac{\hbar}{eB} = 2\pi \left(n + \delta - \beta \right)\]

\noindent  where $A_n$ is the Fermi surface area in the plane perpendicular to the applied magnetic field $B$, $e$ and $\hbar$ are the electron charge and the reduced Planck's constant, respectively, and $n$ is the Landau level (LL) index. $\delta$ is the Maslov index contribution to the phase, equals to $1/2 + \gamma$, where $\gamma$ is 0 and $\pm 1/8$ in two- and three-dimensions, respectively. Finally, $\beta$ is the Berry phase contribution $\Phi_\mathrm{B}$ divided by $2\pi$ and assumes a value between 0 and 1. This LL quantization imparts an oscillatory component to transport and thermodynamic properties, allowing for a direct measurement of $\beta$. Berry phase measurements have attracted considerable attention in the context of monolayer graphene, where a clear signature of a non-zero Berry phase was observed via Shubnikov - de Haas experiments~\cite{zhang_experimental_2005,novoselov_two-dimensional_2005}. The reported phase is close to $\pi$ which matches the value predicted for massless Dirac fermions at the $K$ and $K'$ valleys of graphene~\cite{mikitik_manifestation_1999}. Over the last decade, the Berry phase has also been studied by quantum oscillations (QO) in a number of bulk materials, such as LaRhIn$_5$~\cite{goodrich_magnetization_2002,mikitik_berry_2004}, elemental Bismuth~\cite{zhu_angle-resolved_2011}, and BiTeI~\cite{murakawa_detection_2013}, a bulk Rashba semiconductor. A non-trivial Berry phase is also seen as a key signature of the localized states on the surface of strong topological insulators in three dimensions (see, {\it e.g.}, ref.~\cite{fu_topological_2007,ando_topological_2013}).

\begin{figure}[t]
\includegraphics[scale=0.31]{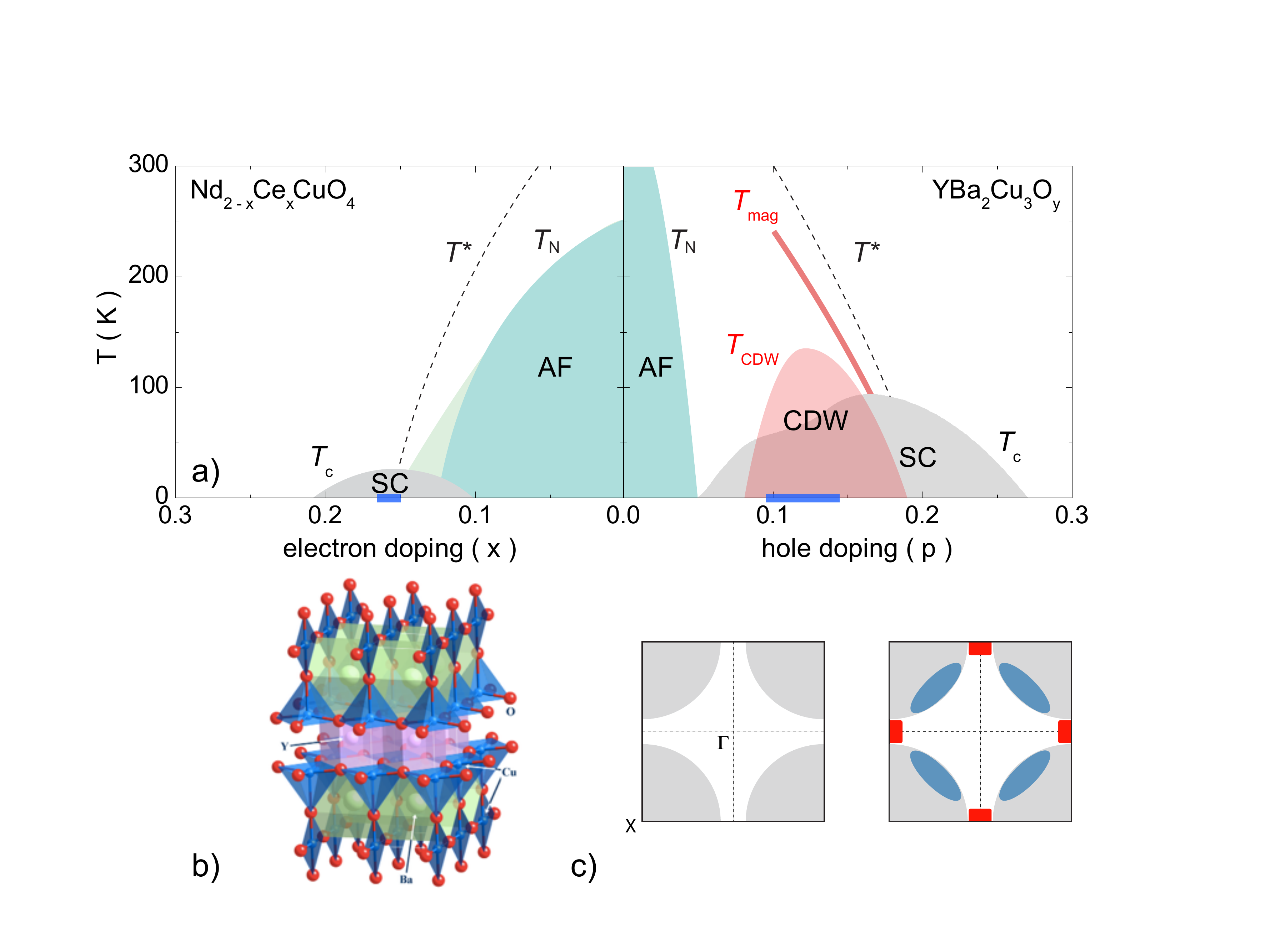}
\caption{{\bf Phase diagram, crystal structure, and Fermi surface of cuprate superconductors}. {\bf a)}: Schematic zero-field temperature-doping phase diagram of cuprate superconductors (adapted from Ref.~\cite{armitage_progress_2010}). At $x$ = $p$ = 0 the parent compound is an antiferromagnetic (AF) Mott insulator. Hole- or electron-doping suppresses the N\'eel temperature $T_N$ and leads to superconductivity (SC) in a dome-shaped region bounded by $T_c$. $T^{\star}$ denotes the pseudogap temperature. Broken time-reversal symmetry is observed below $T_\mathrm{mag}$ by neutron scattering~\cite{fauque_magnetic_2006} and Kerr effect measurements~\cite{xia_polar_2008}. The red region below $T_\mathrm{CDW}$ loosely defines the region where charge density-wave modulations are observed (see main text). The thick blue lines on the doping axes denote the regions where the QO data shown in Fig.~2 were measured. {\bf b)}: Layered crystal structure of YBa$_2$Cu$_3$O$_7$. {\bf c)}: Brillouin zone (square) of the CuO$_2$ plane. The Fermi surface from the band structure (grey) and reconstructed by ($\pi$,$\pi$) antiferromagnetic order, thought to be relevant for electron-doped cuprates, are shown, with the hole and electron pockets in blue and red, respectively.}
\label{Fig1}
\end{figure}

In this Letter we examine the Berry phase in cuprate superconductors. With a layered perovskite crystal structure of alternating CuO$_2$ planes as shown in Fig.~\ref{Fig1}b, the cuprates are highly anisotropic and are essentially quasi-2D materials. At half-filling, the parent compound is an antiferromagnetic Mott insulator and the superconducting phase with record $T_c$ appears with doping, as seen in the simplified phase diagrams in Fig.~\ref{Fig1}a where both the electron- and hole-doped sides are displayed. While the superconducting state is well described by a BCS-like theory with a $d$-wave order parameter, the normal state proves to be more complex with a rich phenomenology and many points of controversy. In particular, the origin of the pseudogap phase, characterized by a strong suppression of the density of states, remains enigmatic, as is the normal-state at high magnetic field and low temperature.

A significant aspect of this normal state was provided by the observation of QO in underdoped YBa$_2$Cu$_3$O$_{y}$ (YBCO)~\cite{doiron-leyraud_quantum_2007}, which gave indications of electronic reconstruction by a broken symmetry state. Since this initial study, QO measurements in cuprates have reached an exquisite level of refinement~\cite{ramshaw_angle_2010,sebastian_quantum_2012} and have been extended to a number of key materials such as overdoped Tl$_2$Ba$_2$CuO$_{6+\delta}$~\cite{vignolle_quantum_2008}, electron-doped Nd$_{2-x}$Ce$_x$CuO$_4$ (NCCO)~\cite{helm_evolution_2009}, and underdoped HgBa$_2$CuO$_{4 + \delta}$ (Hg1201)~\cite{barisic_universal_2013}. The observation of QO is generally seen as an indication of a Fermi liquid and thus far QO data on cuprates show no significant deviation from Lifshitz-Kosevitch (LK) behavior~\cite{sebastian_fermi-liquid_2010}.

While charge density modulations were observed over the last decade in some cuprate materials by scanning tunneling microscopy experiments~\cite{kohsaka_intrinsic_2007}, it was only recently observed in underdoped YBCO by NMR~\cite{wu_magnetic-field-induced_2011} and X-ray diffraction~\cite{ghiringhelli_long-range_2012,chang_direct_2012}, and in underdoped Hg1201 by transport~\cite{barisic_universal_2013} and X-ray diffraction~\cite{tabis_connection_2014}, providing strong evidence that the charge density wave (CDW) is a universal property of hole-doped cuprates. The microscopic origin and nature of the CDW order, and its relation to the pseudogap state and high-temperature superconductivity are currently at the center of attention.

\begin{figure}[t!]
\includegraphics[scale=1.48]{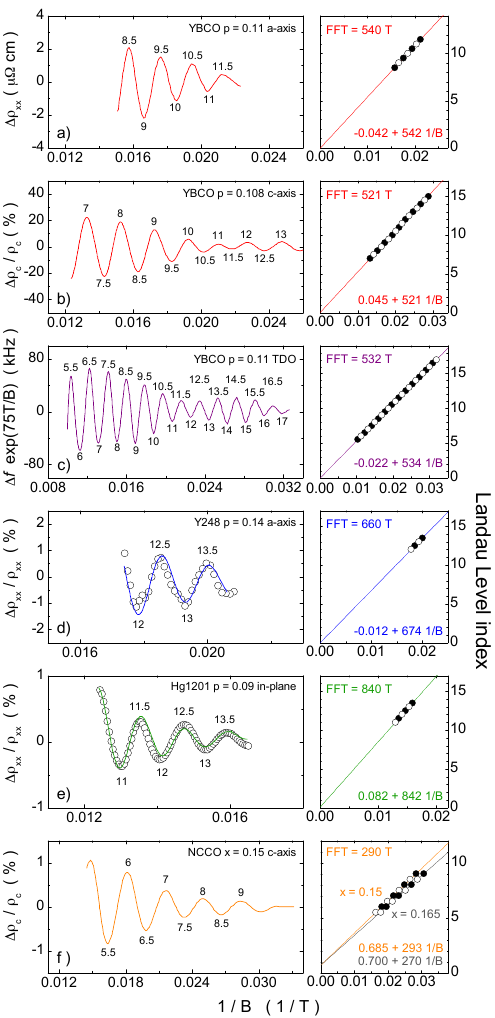}
\caption{{\bf Landau level fan diagram}. {\it Left column}: Quantum oscillatory part $\Delta\rho$ of the electrical resistivity as a function of inverse magnetic field $B$ in: a) YBCO ($p$ = 0.11; this work) b) YBCO ($p$ = 0.108; ref.~\cite{vignolle_quantum_2013}), c) YBCO ($p \approx 0.11$; ref.~\cite{sebastian_quantum_2012}), d) Y248 ($p$ = 0.14; ref.~\cite{bangura_small_2008}), e) Hg1201 ($p \approx 0.09$; ref.~\cite{barisic_universal_2013}), and f) NCCO ($x$ = 0.15 and 0.165; refs.~\cite{kartsovnik_fermi_2011, helm_correlation_2014}). In all panels the lines are measured data, except for Y248 and Hg1201 (panels d) and e))  where the data are circles and the line is a fit to the data, as reported. In panel c) we show the shift in resonance frequency of a tunnel diode oscillator (TDO), proportional to the in-plane conductivity. In all panels the magnetic field is along the $c$-axis. The electrical current is applied as indicated. The numbers are the Landau level index (see main text). {\it Right column}: Landau level index $n$ versus $1/B$. The lines are free linear fits to the data, whose equations are indicated. The Landau levels are determined by ensuring that $\beta$ is between 0 and 1. For NCCO, we also show $n$ vs $1/B$ for $x = 0.165$ (dark red; see supplementary Fig.~3), which has a similar intercept.}
\label{Fig2}
\end{figure}

In Fig.~\ref{Fig2} we show published and unpublished quantum oscillations data on cuprates. From top to bottom, we show data for $a$-axis resistivity in YBCO ($p$ = 0.11), $c$-axis resistivity in YBCO ($p$ = 0.108)~\cite{vignolle_quantum_2013}, in-plane conductivity on YBCO ($p \approx 0.11$) as measured by the shift in resonance frequency of a tunnel diode oscillator (TDO)~\cite{sebastian_quantum_2012}, $a$-axis resistivity on stoichiometric Y248 ($p$ = 0.14)~\cite{bangura_small_2008}, in-plane resistivity on Hg1201 ($p \approx 0.09$)~\cite{barisic_universal_2013}, and $c$-axis resistivity on the electron-doped cuprate NCCO ($x$ = 0.15 and 0.165)~\cite{helm_correlation_2014}. All displayed data correspond to the oscillatory part of the signal plotted as a function of inverse applied field $1/B$, with the field applied along the $c$-axis.

Shubnikov - de Haas (SdH) oscillations of the in-plane conductivity $\sigma_{xx}$ versus the inverse magnetic field $1/B$ are described by an oscillatory component,

\[\Delta\sigma_{xx} \propto \mathrm{cos} \left[ 2\pi \left(\frac{B_{f}}{B} - \delta + \beta  \right) \right] \]

\noindent where $B_{f}$ is the frequency that is proportional to the Fermi surface area. In our analysis, we assume the two-dimensional value $\delta = 1/2$ as justified by the strong transport anisotropy between the plane and the $c$-axis~\cite{hussey_angular_1996,armitage_progress_2010}. By comparison, BiTeI is far less anisotropic than the cuprates, and yet $\delta = 1/2$ is observed~\cite{murakawa_detection_2013}. The conductivity $\sigma_{xx}$ is at a minimum when the chemical potential is halfway between two Landau levels, such that an integer number $n$ of levels are filled. Maxima in $\sigma_{xx}$ correspond to half-integer index. The $n$-th minimum therefore occurs when the argument of the cosine equals $\pi (2n -1)$, thus a plot of $n$ versus $1/B$ gives a straight line whose slope is $B_f$ and whose $y$-intercept is the Berry phase contribution $\beta = \Phi_\mathrm{B}/2\pi$.

%
\begin{figure}[t!]
\includegraphics[scale=0.42]{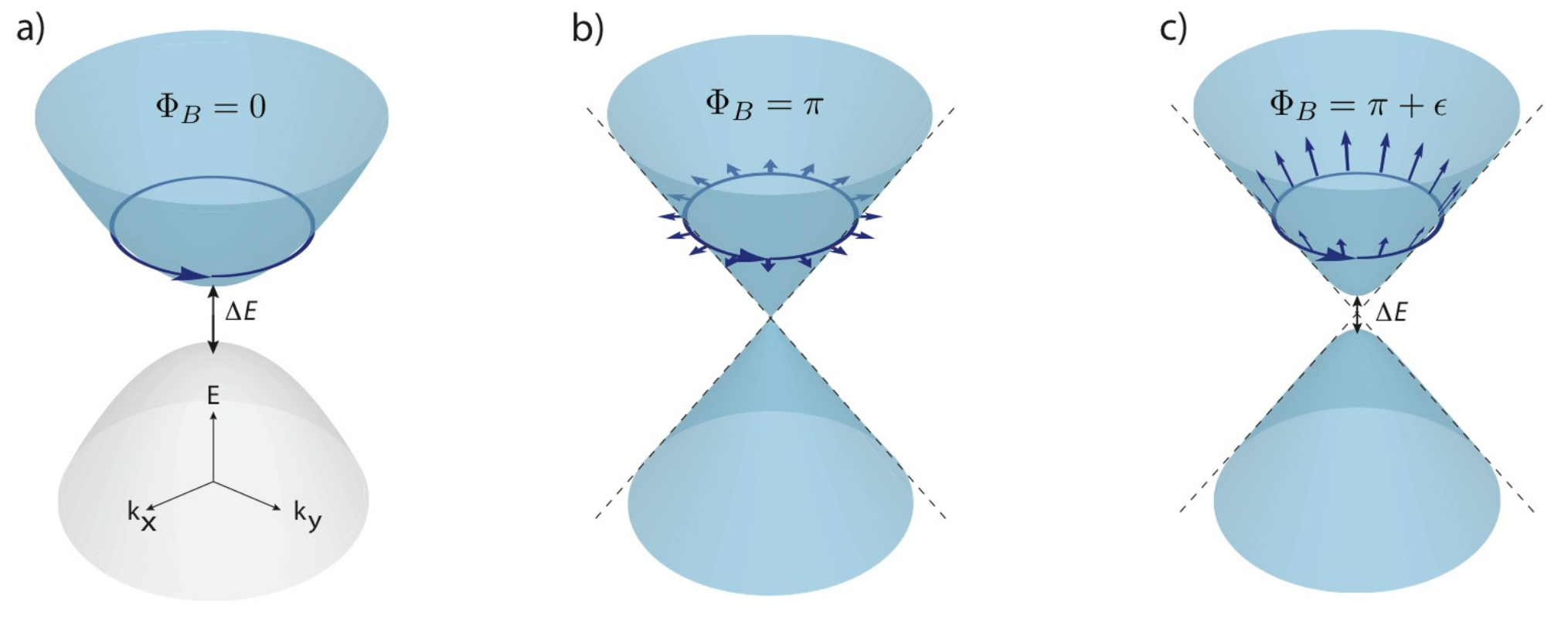}
\caption{{\bf Illustration of Berry phase accumulation in momentum space}. a) Parabolic dispersion of two uncoupled, gapped bands with a trivial phase accumulation of zero mod($2\pi$). Two coupled, linear bands with (b) gapless and (c) gapped dispersion and a phase accumulation of $\simeq \pi$. Note that it is the topological Berry phase of coupled bands that can produce a phase mismatch of $\simeq \pi$~\cite{fuchs_topological_2010,goerbig_measure_2014}.}
\label{Fig3}
\end{figure}
%

When both the magnetic field and electrical current are along the $c$-axis $\sigma_c = 1 / \rho_c$, so that an integer level filling corresponds to a minimum in $\sigma_c$ and thus a maximum in $\rho_c$. For in-plane transport and assuming tetragonal symmetry, the conductivity tensor takes the form $\sigma_{xx} =  \rho_{xx} / (\rho_{xx}^2 + \rho_{xy}^2)$, with $\rho_{xx}$ and $\rho_{xy}$  the longitudinal and transverse resistivity, respectively. In the quantum Hall regime of 2D electron gases, the condition $\rho_{xy} \gg \rho_{xx}$ often holds implying  $\sigma_{xx} \simeq \rho_{xx} / \rho_{xy}^2$, and so the minima of $\rho_{xx}$ and $\sigma_{xx}$ occur simultaneously. This condition also holds in underdoped cuprates where the small carrier density provides a large Hall response.  In supplementary Fig. 1 we show the reconstructed $\sigma_{xx}$ for YBCO $p = 0.11$ and compare its oscillatory part $\Delta\sigma_{xx}$ to that of $\rho_{xx}$: the oscillations are very well matched, with no observable phase shift between the two quantities. We therefore apply the LL indexing of $\sigma_{xx}$ to the in-plane resistivity data in YBCO, Y248, and Hg1201, and plot in Fig.~\ref{Fig2} $n$ versus $1/B$, to which we apply a {\it free linear fit}. In all cases the slope matches well the main QO frequency $B_f$ determined from the fast Fourier transform (FFT), listed in Fig.~\ref{Fig2}. Moreover, the residuals, shown in supplementary Fig.~2, confirm that the linear fits to the data are reliable. From the $y$-intercept of the fits we extract the Berry phase contribution, $\beta$. As seen in Fig.~\ref{Fig2}, a clear and consistent picture emerges from our analysis: to within error bars (see supplementary Fig.~2), the high-field normal state of hole-doped cuprates has a trivial Berry phase of zero mod($2\pi$) whereas the electron-doped NCCO shows a non-trivial phase accumulation $\beta= 0.685\pm 0.031$.

These findings of the Berry phase have profound implications as they set a constraint for theoretical descriptions of the high-field normal state of cuprates, in particular for the case of hole-doped material where no Berry phase accumulation is observed. In general, a non-trivial Berry phase occurs from the crossing of  two bands in the electronic spectrum such as for the (massless) Dirac dispersion found in graphene \cite{zhang_experimental_2005,novoselov_two-dimensional_2005} (see Fig.~\ref{Fig3}b). In a minimal description of the system in the vicinity of the band crossing one can write the Hamiltonian as a matrix in the two dimensional band index space. If the bands disperse linearly around the crossing point, then the matrix takes the form of a Dirac Hamiltonian $ H_k = \vec k \cdot \vec\sigma$ where $\vec\sigma$ is a vector of Pauli matrices that act in band space.

The winding of momentum around contours of constant energy causes a winding of the band pseudospin since it is locked to the momentum. This leads to a $\Phi_\mathrm{B}=\pi$ contribution to the Onsager relation.  While this is not surprising in the context of a Dirac point, it has been predicted that the same Berry phase contribution accumulates around a massive (gapped) Dirac point as long as the Dirac mass, related to the the energy gap in the spectrum, is much smaller than the quasiparticle mass~\cite{fuchs_topological_2010}(see Fig.~\ref{Fig3}c).  On the other hand, as the Dirac mass, and hence the gap, becomes large the Berry contribution reduces and eventually vanishes~\cite{goerbig_measure_2014}. The above description also applies to spin orbit coupled systems in which electron spin rather than pseudo spin, is locked to the momentum direction.  This leads to a $\pi$ Berry phase as in the case of surface states in strong three dimensional topological insulators.

The absence of a Berry phase contribution in hole-doped cuprates points to a normal state with small Fermi pockets composed of decoupled bands.  As far as the QO are concerned, the quasiparticles in these bands move on cyclotron orbits free of any Berry phase accumulation, much like quasiparticles in the Fermi liquid state in metals and semiconductors. While prior QO measurements are consistent with a Fermi liquid state~\cite{sebastian_fermi-liquid_2010}, the absence of a Berry phase contribution reported here strengthen this claim and puts strong constraints on the theoretical description of the underlying electronic structure. For instance, any scenario which involves a $d$-wave like order parameter, whether a precursor of superconductivity or related to another mechanism, with nodes on the Fermi surface leads to Dirac points~\cite{melikyan_quantum_2008,banerjee_theory_2013}. These points should exhibit a $\pi$ Berry phase ($\beta = 1/2$) which is not supported by our findings.

Another interesting point is the observation of the Kerr effect in hole doped compounds~\cite{xia_polar_2008,he_single-band_2011}. The non-zero Kerr rotation may be interpreted as resulting from time reversal symmetry breaking, which implies a non-zero Berry phase is expected~\cite{orenstein_berry_2013}.  This seems to be at odds with our findings, especially in YBCO where both Kerr~\cite{xia_polar_2008} and quantum oscillations measurements are available.

On the electron-doped side of the phase diagram, NCCO shows a phase mismatch of $\beta = 0.685\pm 0.031$ that can only be explained by Berry phase accumulation given that its value is non-trivial and well beyond our error bars. This could be related to the band coupling mechanisms discussed above. Moreover, Fermi surface reconstruction is most likely very different between electron- and hole-doped cuprates, with commensurate antiferromagnetic and bi-axial charge density-wave order thought to be the dominant mechanisms, respectively. It is therefore possible that band coupling such as that shown in Fig.~\ref{Fig3}c occurs in NCCO but not in YBCO. This scenario is supported by the observation of simultaneous breakdown orbits in the QO spectrum~\cite{helm_correlation_2014}, showing that the gap between different reconstructed bands is small. We also note that the magnetism and/or spin-orbit coupling associated with Nd in NCCO might provide another route for phase accumulation.

Our analysis of Berry phase measurements, which can be loosely thought of as interferometry experiments in the Brillouin zone, reveals that the normal state of hole-doped cuprates, which bears the seeds of high-temperature superconductivity, has a trivial Berry phase of zero mod($2\pi$). While in agreement with the expected behaviour of a conventional degenerate Fermi liquid metal, this Berry phase provides new and significant constraints on the underlying nature of cuprates that theoretical descriptions must take into account.\\

{\bf Acknowledgements}\\
We thank W. P. Halperin, L. Taillefer, and A.-M. S. Tremblay for discussions and comments. We thank S. Gaucher for assistance with the manuscript. This work was funded by  Natural Sciences and Engineering Research Council of Canada (NSERC), the Fonds Qu\'eb\'ecois de la Recherche sur la Nature et les Technologies FRQNT (Qu\'ebec), the Canadian Institute for Advanced Research (CIFAR) and the Canada Research Chairs program. A portion of this work was performed at the LNCMI-CNRS in Toulouse, member of the European Magnetic Field Laboratory (EMFL). All (new) data and analysis details presented in this work are available upon request to G.G..

\clearpage

\setcounter{figure}{0}

\section{Supplementary Information for Berry phase in cuprate superconductors}

\subsection{Data and references}

All the quantum oscillation data presented here were measured at low temperature with the magnetic field oriented along the crystalline $c$-axis. The $a$-axis resistivity data on YBCO 6.54 shown in the top panel of Fig.~2 were measured at the LNCMI in Toulouse using a method described in Ref.~\cite{doiron-leyraud_quantum_2007}. All the other data discussed in the present paper are published data and were extracted by digitization. We list in supplementary Table~\ref{tab1} below the details of each data set and indicate the source reference.

{\footnotesize
\begin{table*}
\renewcommand{\tablename}{SUPP. TABLE}
  \centering
\begin{tabular}{| c | c | c |}
\hline
{\bf Material} & {\bf Orientation} & {\bf Reference} \\
\hline
\hline
YBa$_2$Cu$_3$O$_{6.54}$ $p$ = 0.11 & $a$-axis & unpublished \\
\hline
YBa$_2$Cu$_3$O$_{6.59}$ $p$ = 0.108 & $c$-axis & B. Vignolle {\it et al.}, Comptes Rendus Physique {\bf 14}, 39 (2013) \\
\hline
YBa$_2$Cu$_3$O$_{6.56}$ $p  \approx  0.11$ & TDO (in-plane) & S. Sebastian {\it et al.}, Phys. Rev. Lett. {\bf 108}, 196403 (2012) \\
\hline
YBa$_2$Cu$_4$O$_8$ $p$ = 0.14 & $a$-axis & A. F. Bangura {\it et al.}, Phys. Rev. Lett. {\bf 100}, 047004 (2008) \\
\hline
HgBa$_2$CuO$_{4 + \delta}$ $p \approx 0.09$ & in-plane & N. Bari{\v s}i\'c {\it et al.}, Nature Phys. {\bf 9}, 761 (2013) \\
\hline
Nd$_{1.85}$Ce$_{0.15}$CuO$_4$ & $c$-axis & T. Helm {\it et al.}, arXiv:1403.7398 (2014) \\
\hline
Nd$_{1.835}$Ce$_{0.165}$CuO$_4$ & $c$-axis & M. V. Kartsovnik {\it et al.}, New J. Phys. {\bf 13}, 015001 (2011) \\
\hline
\end{tabular}
\caption{Materials ($p$ is the actual carrier concentration per planar Cu atom), electrical current orientation, and source references for the data shown in Fig.~2 of the main text.}
\label{tab1}
\end{table*}
}

\subsection{Comparison of quantum oscillations in $\rho_{xx}$ and $\sigma_{xx}$}

In Shubnikov - de Haas oscillations the Landau levels index are well defined in the conductivity $\sigma_{xx}$: minima and maxima correspond to integer and half-integer index, respectively. In solids, the conductivity tensor is the inverse of the resistivity tensor and for a tetragonal system the diagonal part is expressed as:

\[\sigma_{xx} = \frac{\rho_{xx}}{\rho_{xx}^2 + \rho_{xy}^2}\]

\noindent where $\rho_{xx}$ and $\rho_{xy}$ are the longitudinal and transverse resistivity, respectively. In underdoped YBCO the small Fermi surface gives a large Hall resistivity such that $\rho_{xy} \gg \rho_{xx}$, implying that $\sigma_{xx} \simeq \rho_{xx} / \rho_{xy}^2$: the level index can therefore be determined directly from the resistivity $\rho_{xx}$. For the YBa$_2$Cu$_3$O$_{6.54}$ sample whose oscillatory resistivity is shown in the top panel of Fig.~2, we have computed the full conductivity tensor using the measured $\rho_{xx}$ and $\rho_{xy}$, and extracted its oscillatory part $\Delta\sigma_{xx}$. As shown in supplementary Fig.~1, the positions of the minima and maxima in $\Delta\rho_{xx}$  and $\Delta\sigma_{xx}$ are identical, showing that our indexing scheme is valid.


\begin{figure}[h!]
\renewcommand{\figurename}{SUPP. FIG.}
\includegraphics[scale=1.5]{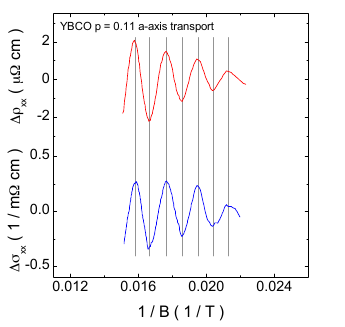}
\caption{Oscillatory part of the electrical resistivity $\Delta\rho_{xx}$ (red) and conductivity $\Delta\sigma_{xx}$ (blue) for YBa$_2$Cu$_3$O$_{6.54}$ at $T$ = 4.2 K as a function of inverse magnetic field $1/B$. The vertical grey lines indicate the positions of the maxima and minima in both quantities, showing that they match well.}
\label{FigS1}
\end{figure}



\begin{figure}[h!]
\renewcommand{\figurename}{SUPP. FIG.}
\includegraphics[scale=1.2]{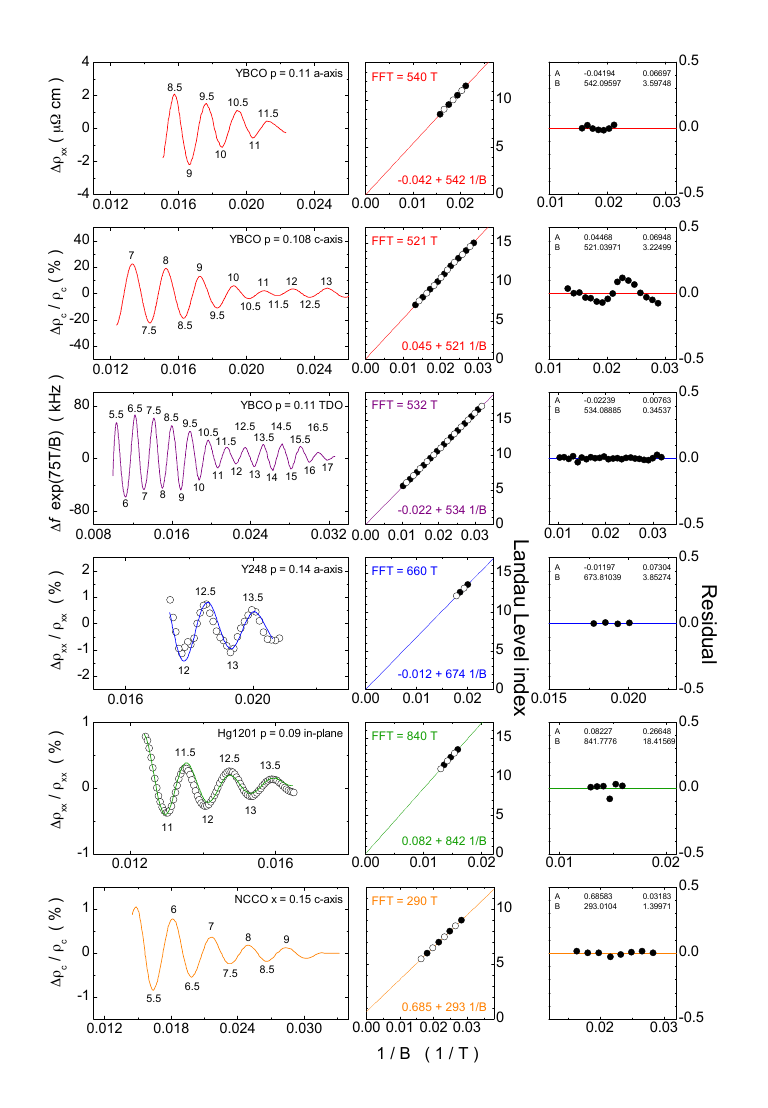}
\caption{{\it Left and middle columns}: Same data as shown in Fig.~2 of the main text. {\it Right column}: Residual difference between the fan diagram data points and the linear fits. The values and errors from the linear regression are indicated in the top of each panel.}
\label{FigS2}
\end{figure}


\subsection{Landau levels fan diagrams and linear fits}

The oscillatory part of the conductivity is expressed as

\[\Delta\sigma_{xx} \propto \mathrm{cos} \left[ 2\pi \left(\frac{B_{f}}{B} - \delta + \beta  \right) \right] \]

\noindent where $B_{f}$ is the frequency proportional to the Fermi surface area $A$, $\beta$ is the Berry phase $\Phi_\mathrm{B}$ divided by $2\pi$, and $\delta$ is the Maslov index contribution to the phase which depends on the dimensionality of the system. It assumes the values $\delta= 1/2$ in 2D and  $\delta = 1/2 \pm 1/8$ in 3D. In our analysis, the two-dimensional value is assumed. Minima in the conductivity occurs when the argument of the cosine is equal to a multiple of $(2n -1)\pi$:

 \[2\pi \left(\frac{B_{f}}{B} - \frac{1}{2} + \beta \right) = (2n -1)\pi\]
 
\noindent It follows that a plot of $n$ vs $1/B$ makes a straight line with a slope of $B_f$ and an intercept at $1/B \rightarrow 0$ equals to $\beta$. To extract the Berry phase, we employ this procedure and make a free linear fit to the fan diagrams. As seen in supplementary Fig.~2, the residuals between our fits and the data is small and shows no systematic deviation. Moreover, the fits give a slope that closely match the frequency extracted from a FFT analysis. Consequently, the values for the Berry phase are reliable. In the right column in supplementary Fig.~2, we indicate the values and errors coming from the linear regression.

\subsection{Berry phase analysis in NCCO}

To ensure that the non-trivial Berry phase in NCCO is a robust result we have analyzed the Berry phase at different dopings. In supplementary Fig.~3 we show our analysis at $x = 0.15$ and 0.165, which reveals in both cases a value for $\beta$ near 0.7.


\begin{figure}[h!]
\renewcommand{\figurename}{SUPP. FIG.}
\includegraphics[scale=1.4]{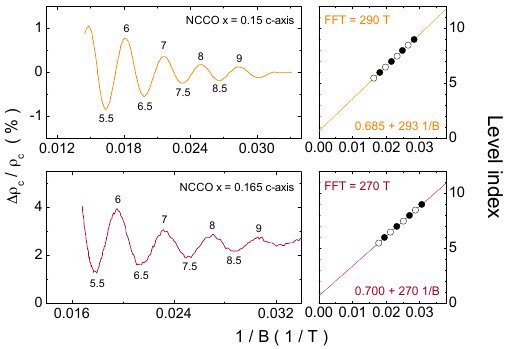}
\caption{{\it Left column}: Quantum oscillatory part of the electrical resistivity as a function of inverse magnetic field for NCCO at $x$ = 0.15 and 0.165 (data from Ref~\cite{kartsovnik_fermi_2011}). The magnetic field and electrical current are along the $c$-axis. The numbers are the Landau level index. {\it Right column}: Landau level index $n$ versus $1/B$. The lines are free linear fits to the data, whose equations are indicated.}
\label{FigS3}
\end{figure}

\end{document}